 \documentstyle[prl,epsfig,aps]{revtex}

\begin{document}
\draft

%
  \wideabs{
%

\title{Franck-Condon-Broadened Angle-Resolved Photoemission Spectra 
Predicted in LaMnO$_3$}

\author{Vasili Perebeinos and Philip B. Allen}
\address{Department of Physics and Astronomy, State University of New York,
Stony Brook, NY 11794-3800}
\date{\today}
\maketitle

\begin{abstract}
The sudden photohole of least energy created in the photoemission process is 
a vibrationally excited state of a small polaron. Therefore the photoemission 
spectrum in LaMnO$_3$ is predicted to have multiple Franck-Condon vibrational 
sidebands. This generates an intrinsic line broadening $\approx$ 0.5 eV.
The photoemission 
spectral function has two peaks whose central energies disperse with band 
width $\approx$ 1.2 eV. Signatures of these 
phenomena are predicted to appear in angle-resolved photoemission spectra. 
\end{abstract}


%
  }   
%

The colossal magnetoresistance (CMR) effect  \cite{Kusters}
in doped manganese oxides 
has attracted a lot 
of attention.
The interplay of charge, orbital and magnetic order  
results in a very rich phase diagram \cite{giant}.   
The parent compound LaMnO$_3$ has 
orthorhombic symmetry at low temperature. 
The Mn$^{+3}$ ion has $d^4$ ($t_{2g}^{3}$,$e_g^{1}$) 
configuration with
an ``inert'' $t_{2g}$ core (spin 3/2) and  
a half-filled doubly degenerate $e_g$-type $d$ 
orbital which is Jahn-Teller (JT) unstable. 
Ignoring rotation of the MnO$_6$ octahedra, which occurs below 1010 K, 
the JT symmetry breaking is cubic to tetragonal \cite{Rodriguez} at 
${\rm T}_{\rm JT}=$750 K.
The corresponding orbital order \cite{Murakami} has 
$x$- and $y$-oriented E$_g$ orbitals alternating in the $x-y$ plane 
with wave vector $\vec{Q}=(\pi,\pi,0)$. This in turn causes layered 
antiferromagnetic (AFA) order to set in at T$_{\rm N}=$140 K.

The electronic structure of LaMnO$_3$ has been studied, for example, 
by photoemission \cite{Park,Dessau} and 
by first principles calculations \cite{Pickett}. Still there is 
controversy about the nature of the low energy excitations,
arising from the interplay between strong on-site Coulomb repulsion 
(which leads to magnetic order) and strong electron-phonon (e-p) interactions 
\cite{Kanamori} (which lead to orbital order).

\begin{figure}
\psfig{figure=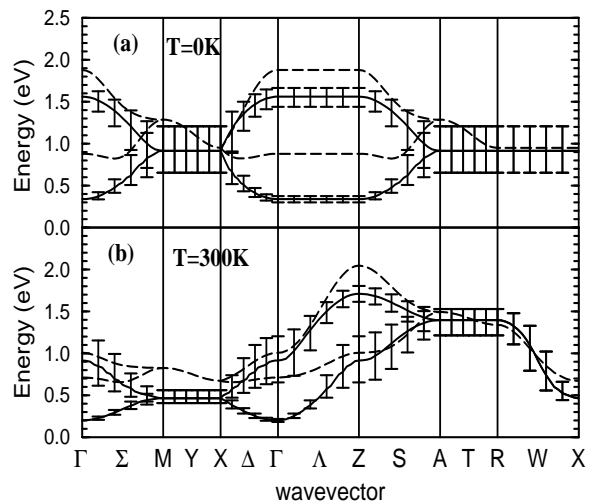,height=2.7in,width=3.0in,angle=0}
\caption{Peak positions of the ARPES (a) at T=0 K 
and (b) at T=300 K along high- 
symmetry lines of the tetragonal Brillouin zone are shown by solid lines. 
Spin disorder at T above the Neel temperature T$_{\rm N}$ effectively reduces 
the hopping parameter t=(dd$\sigma$) by 2/3 and adds hopping in $\hat{z}$ 
direction which strongly affects dispersion. Error bars 
(FWHM of the photoemission peak) represent Franck-Condon broadening. 
Band structure (U=0 limit) peak  positions $\lambda_{1,2}(\vec{k})$ 
are shown by dashed lines.} 
\label{fig1}
\end{figure}

When an electron is removed from the JT-ordered 
ground state, e-p coupling causes the hole to self-localize in an 
``anti-JT'' small polaron state. In a previous paper \cite{Allen1} we 
have described the localized polaron in adiabatic approximation. Residual 
non-adiabatic coupling allows the hole to disperse with band width
narrowed by Huang-Rhys factor $e^{-3\Delta/4\hbar\omega}\approx 10^{-4}$. 
The photoemission process is sudden. The emitted electron with wavevector 
$\vec{k}$ leaves a hole in a lattice ``frozen'' in the unrelaxed JT state.
Ignoring lattice relaxation, this hole would disperse with band width 
$2t\approx$1 eV ($t$ is the hopping parameter), as shown on Fig. \ref{fig1}. 
However, this is not a stationary state and must be regarded as a 
superposition of exponentially narrowed small polaron bands. Such bands have 
anti-JT oxygen distortions at each site, but a sufficient number of 
vibrational quanta are also excited such that the anti-JT distortion at 
time zero is ``undone''. This is ``Franck-Condon principle''.

The measured spectrum at wavevector $\vec{k}$ will consist of a central 
$\delta$-function at the energy of the frozen lattice (dispersive) band 
$\varepsilon_{1,2}(\vec{k})$, plus multiple vibrational side-bands at energy 
$\varepsilon_{1,2}(\vec{k})\pm n\hbar\omega$, with an overall Gaussian 
envelope whose width is approximately the polaron binding energy.
Franck-Condon broadening has been seen in photoemission spectra of solid 
nitrogen and oxygen  \cite{Himpsel}. 
We find  at each 
wavevector $\vec{k}$ the photoemission spectrum 
has an intrinsic Franck-Condon broadening indicated by error 
bars in Fig. \ref{fig1}. 
The position of the maximum disperses with $\vec{k}$-vector close to the 
``frozen'' lattice spectrum.
A qualitative picture of this process has been given by Sawatzky 
\cite{Sawatzky} in the context of high temperature superconductors and by 
Dessau and Shen \cite{Dessau} for the manganites. 
The present paper gives an exact algebraic prediction for the 
Angle-Resolved-Photoemission-Spectra (ARPES) of a model Hamiltonian 
for LaMnO$_3$.

Our model Hamiltonian \cite{Allen1}, first introduced by Millis 
\cite{Millis}, has hopping ${\cal H}_{\rm el}$, 
electron-phonon  ${\cal H}_{\rm ep}$, and lattice ${\cal H}_{\rm L}$ energies:
\begin{eqnarray}
{\cal H}_{\rm el}&=&t \sum_{\ell,\pm} \left\{[c^{\dagger}_x(\ell)
c_x(\ell \pm \hat{x})] + [x\rightarrow y] +
[y \rightarrow z] \right\}
\nonumber\\
{\cal H}_{\rm ep}&=&-g\sum_{\ell,\alpha} \hat{n}_{\ell,\alpha}
(u_{\ell,\alpha}-u_{\ell,-\alpha})
\nonumber\\
{\cal H}_{\rm L}&=&\sum_{\ell,\alpha}(P^2_{\ell,\alpha}/2M
+Ku^2_{\ell,\alpha}/2).
\label{hmodel}
\end{eqnarray}
In these formulas $c^{\dagger}_{\alpha}(\ell)$ creates a state with orbital
$\psi_{\alpha}=|3\alpha^2-r^2>$, where $\alpha=x,y,z$. These three orbitals 
span the two dimensional $e_g$ subspace and can be expressed in terms of the 
conventional orthogonal basis $\Psi_2=d_{x^2-y^2}$, 
$\Psi_3=d_{3z^2-r^2}=\Psi_z$; specifically 
$\Psi_{x,y}=\pm\sqrt{3}/2\Psi_2-\Psi_3/2$.
The resulting ${\cal H}_{\rm el}$ coincides with  the 
nearest-neighbor two-center Slater-Koster \cite{Slater} hopping Hamiltonian 
with overlap integral $t=(dd\sigma)$ and $(dd\delta)=0$. The hopping 
parameter $t=0.5$ eV is chosen to agree with an {\sl ab initio} 
$e_g$ band width of 1 eV \cite{Pickett}.
The e-p interaction ${\cal H}_{\rm ep}$ is modeled by a linear energy 
reduction of an occupied $\psi_x$ orbital 
($\hat{n}_{\ell,x}=c^{\dagger}_{x}(\ell)c_{x}(\ell)$) 
if the corresponding two oxygens in the $\pm \hat{x}$ direction
expand outwards, and similarly for $\hat{y}$ and $\hat{z}$
oxygens if $\psi_y$ or $\psi_z$ orbitals are occupied.
The strength of the e-p coupling $g$ determines the JT splitting 
$2\Delta=1.9$ eV, which is fitted to agree with the lowest optical 
conductivity peak \cite{Jung,Allen2}. Static oxygen distortions 
$2u=\sqrt{2\Delta/M\omega^2}$=0.296 {\AA}   given by our model agree well 
with neutron diffraction data 0.271 {\AA} \cite{Rodriguez}.
For the lattice term ${\cal H}_{\rm L}$ we use a simplified model where 
oxygen vibrations along Mn-O-Mn bonds are local Einstein oscillators.
The displacement $u_{\ell,\alpha}$ is  measured from cubic 
perovskite position of the nearest oxygen in the $\hat{\alpha}$-direction
to the Mn atom at $\ell$. The 
oxygen vibrational energy  $\hbar\omega=\hbar\sqrt{K/M}=0.075$ eV is taken 
from Raman data \cite{Raman}. 
In addition there is a large on-site Coulomb repulsion U and a large Hund 
energy. These terms inhibit hopping except to empty sites where the t$_{2g}$ 
core spins are aligned correctly.

In adiabatic approximation one can solve this problem for U=0 or U=$\infty$. 
Both cases give a good description of the observed cooperative JT order. 
When U=0, the ground state wavefunction is:
\begin{equation}
|{\rm GS},0>=\prod_{\vec{k}}c_{\vec{k}1}^{\dagger}
                   c_{\vec{k}2}^{\dagger}|{\rm vac}>.
\label{gs0}
\end{equation}
A JT gap $\approx 2\Delta$ opens and the lower two bands of energy 
$\lambda_{\vec{k}1}$, $\lambda_{\vec{k}2}$ are filled. The photohole 
as initially created has energy:
\begin{eqnarray}
\lambda_{1,2}^2=\Delta^2+t^2(2C_x^2+C_xC_y+2C_y^2)\pm t|C_x+C_y|
\nonumber\\
\sqrt{\Delta^2+4t^2(C_x^2-C_xC_y+C_y^2)},
\label{bs}
\end{eqnarray}
where $C_{x,y}=\cos k_{x,y}$ and $C_z$ not entering at T=0 K. These bands 
are shown in Fig. \ref{fig1}(a) as dashed lines.

At this point, at least in principle, could proceed numerically to find the 
polaronic energy lowering. However, it is both easier and more realistic to 
switch to U=$\infty$. The ground state wavefunction:
\begin{equation}
|{\rm GS}, \infty>=\prod_{\ell}^A c_X^{\dagger}(\ell)
    \prod_{\ell^{\prime}}^B c_Y^{\dagger}(\ell^{\prime})|{\rm vac}>,
\label{gs1}
\end{equation}
has orbitals $\Psi_{X,Y}=-(\Psi_3\mp\Psi_2)/\sqrt{2}$ occupied singly on 
interpenetrating $A$ and $B$ sublattices. 
This is a fully correlated 
state with zero double occupancy, while Eq. (\ref{gs0}) is a 
band wavefunction in which two electrons are found on the same Mn atom 
with non-zero probability. For U$\geq$ 6t $\approx 3$ eV, the state (\ref{gs1}) 
has lower energy than state (\ref{gs0}) \cite{Allen1}.
Neglecting creation of orbital defects with energy $\approx 2\Delta$ 
the ``frozen'' lattice approximation predicts photohole energies  
$\Delta+\varepsilon_{1,2}(\vec{k})$, where 
$\varepsilon_{1,2}(\vec{k})=\pm(t/2)(C_x+C_y)$.
We need to add a non-adiabatic treatment of the e-p coupling. 
The  effective Hamiltonian 
${\cal H}_{\rm eff}={\cal H}_{\rm el}+{\cal H}_{\rm ep}+{\cal H}_{\rm L}$
for the single hole is:
\begin{eqnarray}
{\cal H}^{\rm A}_{\rm el}&=&\sum_{\ell \in A}\frac{t}{4}
\bigl(d_Y^{\dagger}(\ell \pm x)d_X(\ell)+
d_Y^{\dagger}(\ell \pm y)d_X(\ell)\bigr)
\nonumber \\
{\cal H}^{\rm A}_{\rm ep}+{\cal H}^{\rm A}_{\rm L}&=&\sum_{\ell \in A}
d_X^{\dagger}(\ell)d_X(\ell)
\biggl[\Delta+\sum_{\alpha}\kappa_{\alpha}
\Bigl(a_{\alpha}(\ell)+
\nonumber\\
a_{\alpha}^{\dagger}(\ell)-b_{\alpha}(&\ell-&\alpha)
-b_{\alpha}^{\dagger}(\ell-\alpha)\Bigr)\biggr]
+
\sum_{\alpha}a_{\alpha}^{\dagger}(\ell)a_{\alpha}(\ell).
\label{eff}
\end{eqnarray}
Here the operator $d^{\dagger}_X(\ell)=c_X(\ell)$ creates a hole in the 
JT ground state by destroying an electron on orbital $X$ at site 
$\ell$ (if $\ell \in A$ sublattice), and the operator  
$d^{\dagger}_Y(\ell)=c_Y(\ell)$ creates a hole on $B$ sublattice 
(if $\ell \in B$). The phonon operators $a_{\alpha}^{\dagger}(\ell)$ or 
$b_{\alpha}^{\dagger}(\ell)$ create vibrational quanta on the 
$\ell+\hat{\alpha}/2$ oxygen, if  $\ell \in A$
or $\ell \in B$ respectively. The e-p coupling 
constants  are
$\kappa_{x,y,z}=\sqrt{\Delta/12}(1+\sqrt{3}/2 ; 1-\sqrt{3}/2 ; 1)$.
The  Hamiltonian and all other energy
parameters $\Delta$ and $t$ in Eq. \ref{eff} are in units of 
$\hbar\omega$. The total Hamiltonian has an additional term 
${\cal H}_{\rm el}^{\rm B}+{\cal H}_{\rm ep}^{\rm B}+{\cal H}_{\rm L}^{\rm B}$
which is obtained from Eq. (\ref{eff}) by interchanging operators:
\begin{equation} 
d_{Y}\leftrightarrow d_{X}, a_{x}\leftrightarrow b_{y},
a_{y}\leftrightarrow b_{x}, a_{z}\leftrightarrow b_{z} 
\label{change}
\end{equation}
and summing over the $B$  sublattice.

\begin{figure}
\centerline{
\psfig{figure=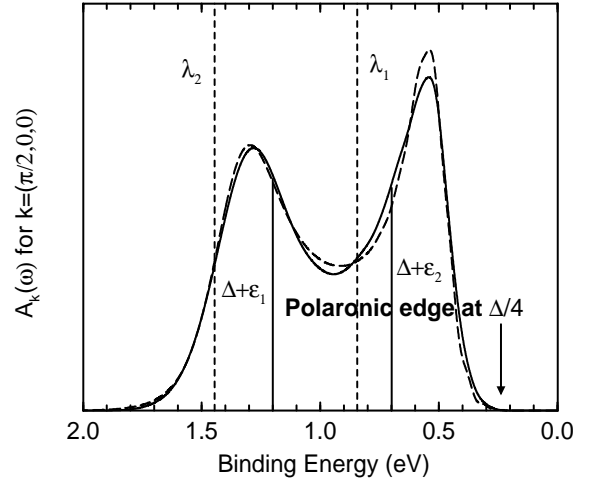,height=2.55in,width=2.95in,angle=0}}
\caption{Energy dependence of the 
imaginary part of the Green function at T=0 and $\vec{k}=(\pi/2,0,0)$. 
The solid line is numerical diagonalization of (\protect\ref{eff}) in the 
subspace (\protect\ref{trial}). The 
long-dashed 
line is the analytical approximation (\protect\ref{eigsol}) and  
(\protect\ref{arpes}). 
The spectral function consists of two asymmetric peaks with  mean values
$\Delta+\varepsilon_{1,2}$.
The zero-phonon 
line is seen at the adiabatic ground state polaron energy $\Delta/4$ 
\protect\cite{Allen1}. The dashed lines show the peak positions $\lambda_{1,2}$ 
of the uncorrelated electron theory Eq. (\protect\ref{bs}). }
\label{fig2}
\end{figure}

Following Cho and 
Toyozawa \cite{Cho} we are able to diagonalize Hamiltonian (\ref{eff}) 
in a very large truncated basis of functions with a hole present 
on site $\ell$ and an arbitrary number of vibrational quanta 
${\rm p}_{\pm x},{\rm p}_{\pm y},{\rm p}_{\pm z}$ on the six 
displaced neighboring oxygens:
\begin{eqnarray}
|\Psi^{\rm A}(\ell,\{p\})>=d_{X}^{\dagger}(\ell)
\prod_{\alpha} {\rm U}_{\ell}^{a_{\alpha}}(-\kappa_{\alpha})
\frac{(a^{\dagger}_{\alpha}(\ell))^{p_{+\alpha}}}{\sqrt{p_{+\alpha}!}}
\nonumber \\
{\rm U}_{\ell-\alpha}^{b_{\alpha}}(\kappa_{\alpha})
\frac{(b^{\dagger}_{\alpha}(\ell-\alpha))^{p_{-\alpha}}}{\sqrt{p_{-\alpha}!}}
|{\rm GS},\infty>.
\label{trial}
\end{eqnarray}
The displacement operator 
${\rm U}_{\ell}^{a}(\kappa)=\exp{[-\kappa(a_{\ell}-a^{\dagger}_{\ell})]}$ 
makes the 
${\cal H}_{\rm ep}+{\cal H}_{\rm L}$ part of the Hamiltonian diagonal.
To get basis functions $|\Psi^{\rm B}(\ell,\{p\})>$ for holes on the $B$ 
sublattice, the operators in Eq. (\ref{trial}) should be 
interchanged according to Eq. (\ref{change}). 
The next step is to build Bloch wavefunctions by Fourier transformation 
of the basis functions Eq. (\ref{trial}).
Then the Hamiltonian (\ref{eff}) will be diagonal with respect to 
$\vec{k}$-vector. The hopping term of the Hamiltonian ${\cal H}_{\rm el}$ 
couples the  $|\Psi^{\rm A}(\ell,\{p\})>$ and 
$|\Psi^{\rm B}(\ell',\{p\})>$-wavefunctions on the neighboring sites.
The vibrational wavefunctions give a product of 
Huang-Rhys factors, but a shared oxygen contributes a non-factorizable 
overlap integral. However if one treats this shared oxygen as two independent 
atoms, one coupled to each site, then the Hamiltonian has a simple form:
\begin{eqnarray}
{\cal H}^{\rm AA}_{pp'}(\vec{k})&=&{\cal H}^{\rm BB}_{pp'}(\vec{k})=
\delta_{\{p\}\{p'\}}
\biggl[\frac{\Delta}{4}+\sum_{\alpha}(p_{+\alpha}+p_{-\alpha})\biggr]
\nonumber\\
{\cal H}^{\rm AB}_{pp'}(\vec{k})&=&{\cal H}^{\rm BA}_{pp'}(\vec{k})=
\varepsilon(\vec{k})
\prod_{\alpha}(-1)^{p_{-\alpha}+p'_{-\alpha}}
\nonumber\\
\biggl[
P(p_{+\alpha}&,\kappa_{\alpha}&)P(p_{-\alpha},\kappa_{\alpha})
P(p'_{+\alpha},\kappa_{\alpha})P(p'_{-\alpha},\kappa_{\alpha})
\biggr]^{1/2},
\label{kspace}
\end{eqnarray}
where $P(p,\kappa)=\exp(-\kappa^2)\kappa^{2p}/p!$ is a 
Poisson distribution.
Since off-diagonal terms factorize, the analytical solution is available
in this approximation:
\begin{eqnarray}
&&\Psi_{\lambda}^{1,2}(\vec{k})=\sum_{\{p\}=0}^{\infty}
\prod_{\alpha}(-1)^{p_{-\alpha}}
\biggl[
P(p_{+\alpha},\kappa_{\alpha})P(p_{-\alpha},\kappa_{\alpha})
\biggr]^{1/2}
\nonumber\\
&&\frac{\Psi_{\vec{k}}^{\rm A}(\{p\})\pm\Psi_{\vec{k}}^{\rm B}(\{p\})}
{\sqrt{2G'(x_{\lambda})}
\bigl(\sum_{\alpha^{\prime}}(p_{+\alpha^{\prime}}+p_{-\alpha^{\prime}})
-x_{\lambda}\bigr)}.
\label{funsol}
\end{eqnarray}
The corresponding eigenvalues are:
\begin{eqnarray}
&&E_{\lambda}^{1,2}(\vec{k})=\frac{\Delta}{4}+x_{\lambda}^{1,2}(\vec{k}) ,
\ \ \   
1+\varepsilon_{1,2}(\vec{k})G(x_{\lambda}^{1,2})=0
\nonumber\\
&&G(x_{\lambda})=e^{-3\Delta/4}\sum_{p=0}^{\infty}\frac{(3\Delta/4)^p}{p!}
\frac{1}{p-x_{\lambda}}.
\label{eigsol}
\end{eqnarray}
The $G'(x_{\lambda})$ function in Eq. (\ref{funsol}) is a first derivative of 
$G(x_{\lambda})$ and makes wavefunctions  normalized. 
The correct solution needs a numerical diagonalization of the Hamiltonian which 
explicitly includes vibrational states of the four shared oxygens.
As can be seen on Fig. \ref{fig2} the difference is negligible between a 
typical spectrum obtained 
using approximation (\ref{eigsol}) and correct numerical treatment.
The ground state of the Hamiltonian (\ref{kspace}), with energy 
(from Eq. \ref{eigsol})  
$E_0(\vec{k})=\Delta/4+x_0(\vec{k})$, corresponds to the anti-JT 
polaron. Its effective mass, deduced from $d^2x_0(\vec{k})/d\vec{k}^2$,
provides a realistic alternative (exact for $\Delta\rightarrow 0$ or 
$\infty$) to the available variational 
approaches \cite{Katja} or exact quantum Monte Carlo simulations \cite{Pasha}. 

An ARPES experiment measures the spectral function 
$A(\vec{k},\omega)=-\frac{1}{\pi}G(\vec{k},\omega)$ with momentum $\vec{k}$ 
fully resolved, provided there is no dispersion in the direction 
perpendicular to the surface. Although LaMnO$_3$ is cubic, because of the 
layered AFA magnetic structure, at low temperatures Mn $e_g$ 
electrons are two-dimensional and the  spectrum can be measured:
\begin{equation}
A(\vec{k},\omega)=
\sum_{f}|<f|d^{\dagger}(\vec{k})|{\rm GS},\infty>|^2\delta(E-E_f).
\label{impart}
\end{equation}
The operator $d^{\dagger}(\vec{k})$
excites a hole from the JT ground state. Summation over final eigenstates 
$|f>$ includes summation over branch index $i=1,2$ and number of  phonons 
$\lambda=0,1...\infty$. 
The sequence of delta functions in Eq. (\ref{impart}) should be replaced by  
convolved local densities of phonon states, which we approximate  
by a Gaussian, 
$\delta(E)\rightarrow\exp(-E^2/2\gamma^2)/\sqrt{2\pi}\gamma$.
Substituting solution (\ref{funsol}), (\ref{eigsol}) into equation 
(\ref{impart}), we obtain the spectral function:
\begin{equation}
A(\vec{k},\omega)=
\sum_{\lambda,i}\frac{G^2(x_{\lambda}^i)}{G'(x_{\lambda})}
\delta(E-E_{\lambda}^i).
\label{arpes}
\end{equation}

Equation (\ref{arpes}) along with (\ref{eigsol}) gives the ARPES spectrum 
normalized to $\int d\omega A(\vec{k},\omega)=1$. 
The first energy moment of the spectrum \cite{Cho} 
coincides with the free hole energy calculated in 
the ``frozen'' lattice approximation $\Delta+\varepsilon_{1,2}(\vec{k})$ 
shown on Fig. \ref{fig2}.   
The edge of the spectrum corresponds to polaron creation at energy  
$\approx \Delta/4$. This transition is weaker by 3 orders of magnitude 
than the peak at $\approx\Delta+\varepsilon_{1,2}(\vec{k})$.

At room temperature magnetic order is lost. The paramagnetic state is modeled 
by a mean field approximation, namely scaling the  effective hopping 
integral by 2/3 and allowing hopping in $\pm\hat{z}$ direction.
This modifies the  single particle energy band entering Eq. (\ref{kspace}) to 
$\varepsilon_{1,2}(\vec{k})=t/3\bigr(-2C_z\pm(C_x+C_y)\bigl)$.
But the JT orbital order is not destroyed at T=300 K and Franck-Condon 
broadening is still expected.
When spins are disordered, $\vec{k}$ is not a good quantum number 
and additional broadening is expected.
Only phonon broadening of the ARPES along with peak positions are 
shown on Fig. \ref{fig1}(b).

\begin{figure}
\psfig{figure=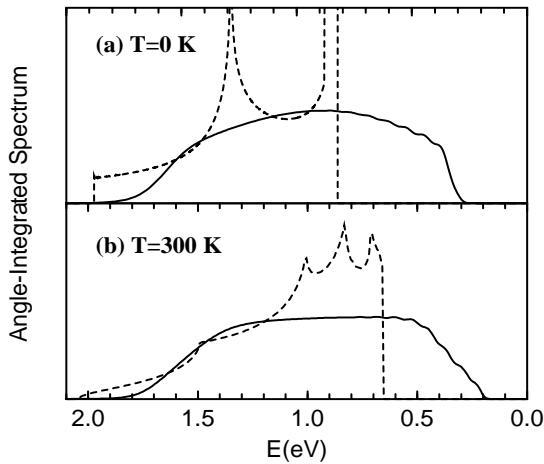,height=2.7in,width=3.20in,angle=0}
\caption{Angle-integrated photoemission spectrum (solid line) at 
(a) T=0 K and (b) ${\rm T}$=300K. At room temperature, spins are 
disordered, but JT order is not destroyed. 
For comparison, the uncorrelated electron (U=0 limit) band 
structure density of states is shown by the dashed lines, with 
2D Van Hove singularities at T=0 K.}
\label{fig3}
\end{figure}

The angle-integrated spectrum, shown on Fig. \ref{fig3} for low and high 
temperatures, has a width of about 1.2 eV and is almost temperature 
independent. The uncorrelated (U=0) band 
structure, shown for comparison, 
is sensitive to  
magnetic order and therefore temperature dependent.

The existing photoemission data \cite{Park,Dessau} are consistent with our 
predictions. Higher resolution experiments are needed to test the theory 
and to unravel the nature of the lowest energy excitations in the LaMnO$_3$. 
To make such an experiment possible, a single domain sample
(having 2D dispersion at T=0K) is needed, with 
good control of  oxygen concentration \cite{Dabrowski}.

\acknowledgements
We thank P. D. Johnson for helpful conversations.
This work was supported in part by NSF Grant No.\ DMR-9725037.


\begin{references}
\bibitem{Kusters}	R. M. Kusters, J. Singelton, D. A. Keen,
			R. McGreevy, and W. Hayes,
			Physica B {\bf 155}, 362 (1989);
         		S. Jin, T. H. Tiefel, M. McCormack, R. A. Fastnacht,
			R. Ramesh, and L. H. Chen,
			Science {\bf 264}, 413 (1994).

\bibitem{giant}		{\sl Colossal Magnetoresistance, Charge Ordering, 
			and Related Properties of Manganese Oxides},
                        edited by C.N.R. Rao and B. Raveau
			(World Scientific, Singapore, 1998).

\bibitem{Rodriguez}	J. Rodriguez-Carvajal, M. Hennion, F. Moussa,
                        A. H. Moudden, L. Pinsard, and A. Revcolevschi,
                        Phys. Rev. B {\bf 57}, 3189 (1998).

\bibitem{Murakami}      Y. Murakami, J. P. Hill, D. Gibbs, M. Blume, I. Koyama,
                        M. Tanaka, H. Kawata, T. Arima, Y. Tokura, K. Hirota,
                        and Y. Endoh, Phys. Rev. Lett. {\bf 81}, 582 (1998).

\bibitem{Park}          J. H. Park, C. T. Chen, S. W. Cheong, W. Bao, 
                        G. Meigs, V. Chakarian and Y. U. Idzerda,  
                        Phys. Rev. Lett. {\bf 76}, 4215 (1996);
                        T. Saitoh A. E. Bocquet, T. Mizokawa, H. Namatame and 
                        A. Fujimori,  Phys. Rev. B {\bf 51}, 13942 (1995).


\bibitem{Dessau}        D. S. Dessau and Z. X. Shen, in 
                        {\sl Colossal Magnetoresistive Oxides},
                        edited by Y. Tokura 
                        (World Scientific, Singapore, 1998).

\bibitem{Pickett}       W. E. Pickett and D. Singh, Phys. Rev. B 
                        {\bf 53}, 1146 (1996); 
                        S. Satpathy, Z. S. Popovi\'{c} and 
                        F. R. Vukajlovi\'{c},
                        Phys. Rev. Lett. {\bf 76}, 960 (1996);
                        I. Solovyev, N. Hamada and K. Terakura, 
                        Rhys. Rev. Lett. {\bf 76}, 4825 (1996); 
                        Y.-S. Su, T. A. Kaplan, 
                        S. D. Mahanti and J. F. Harrison, Phys. Rev. B
                        {\bf 61}, 1324 (2000).

\bibitem{Kanamori} 	J. Kanamori, J. Appl. Phys. {\bf 31}, 14S (1960);
			K. I. Kugel and D. I. Khomskii,
			Sov. Phys. Usp. {\bf 25}, 231 (1982).

\bibitem{Allen1}        P. B. Allen and V. Perebeinos, Phys. Rev. B {\bf 60},
                        10747 (1999).

\bibitem{Himpsel}       F. J. Himpsel, N. Schwentner, E. E. Koch,
                        Phys. Status Solidi B {\bf 71}, 615 (1975). 

\bibitem{Sawatzky}      G. A. Sawatzky, Nature {\bf 342}, 480 (1989).

\bibitem{Millis}	A. J. Millis, Phys. Rev. B {\bf 53}, 8434 (1996).

\bibitem{Slater}	J. C. Slater and G. F. Koster,
			Phys. Rev. {\bf 94}, 1498 (1954).

\bibitem{Jung}		J. H. Jung, K. H. Kim, D. J. Eom, T. W. Noh,
			E. J. Choi, J. Yu, Y. S. Kwon, and Y. Chung,
			Phys. Rev. B {\bf 55}, 15489 (1997);
			J. H. Jung, K. H. Kim, T. W. Noh, E. J. Choi, and 
                        J. Yu,Phys. Rev. B {\bf 57}, 11043 (1998).

\bibitem{Allen2}        P. B. Allen and V. Perebeinos, Phys. Rev. Lett.
                        {\bf 83}, 4828 (1999).

\bibitem{Raman}         M. N. Iliev, M. V. Abrashev, H.-G. Lee,
                        V. N. Popov, Y. Y. Sun, C. Thomsen,
                        R. L. Meng, and C. W. Chu,
                        Phys. Rev. B {\bf 57}, 2872 (1998);
                        V. B. Podobedov, A. Weber, D. Romero,
                        J. P. Rice and H. D. Drew,
                        Phys. Rev. B {\bf 58}, 43 (1998).
 
\bibitem{Cho}		K. Cho and Y. Toyozawa,
			J. Phys. Soc. Jpn. {\bf 30}, 1555 (1971).

\bibitem{Katja}         A. H. Romero, D. W. Brown and K. Lindenberg,
                        Phys. Rev. B {\bf 59}, 13728 (1999).

\bibitem{Pasha}         P. E. Kornilovitch, Phys. Rev. Lett. 
                        {\bf 84}, 1551 (2000).

\bibitem{Dabrowski}     B. Dabrowski, R. Dybzinski, Z. Bukowski, O. Chmaissem,
                        J. D. Jorgensen, J. Solid State Chem. 
                        {\bf 146}, 448 (1999).


\end{references}
\end{document}